# In-situ Synthesis of Bismuth (Bi)/Reduced Graphene Oxide (RGO) Nanocomposites as High Capacity Anode Materials for Mg-ion Battery

Tirupathi Rao Penki[a,*,†], Geetha Valurouthu[b,c,†], S. Shivakumara[a],
Vijay A. Sethuraman[c] and N. Munichandraiah[a,*]

[a]*Department of Inorganic and Physical Chemistry, Indian Institute of Science-Bangalore, India-560012.*
[b]*Bachelor of Science (Research) Programme, Indian Institute of Science-Bangalore, India-560012.*
[c]*Department of Materials Engineering, Indian Institute of Science, Bangalore, India-560012.*

*E-mail: tiru.penki@gmail.com and muni@ipc.iisc.ernet.in
† Equal contributions.

## 1 Introduction

Magnesium (Mg) based batteries have drawn much attention globally, and may be a good alternative for lithium-based batteries in the future, due to their multivalent energy storage system, low electrochemical potential and high volumetric specific capacity.[1-4] The specific volumetric capacity of the Mg anode is 3,833 mAh cc$^{-3}$, which is twice that of Li-metal and five times that of the graphite anode.[2] Mg is a benign element, the 5$^{th}$ most abundant in the earth's crust and more importantly three orders of magnitude more abundant than Li. In perspective of economic merits, magnesium battery has greater potential to become a more cost-friendly technology than the current Li-ion battery for portable, grid and electrical applications. Additionally, Mg has superior thermal and air stability due to the formation of a protective thin oxide film on its surface. Its non-dendritic electrochemical behaviour unlike that of Li, attracts electrochemists to further advance Mg-based batteries.[5] Mg based batteries are so popularized in a short span of time but working on Mg batteries is still not fully exploited due to the lack of appropriate high performance electrode materials and electrolytes, and also due to the inadequate control of the electrochemical reaction at the metal-electrolyte interfaces.[5-12] Mg metal forms a non-conducting passive layer on the surface, which prevents the reversible Mg plating/stripping.[11-14] Specially designed and synthesised electrolytes are needed for Mg metal anode and only limited numbers of electrolytes are reported.[2, 5, 9, 11, 15-17]

The incompatibility between electrolyte and anode is due to the inability to conduct Mg$^{2+}$ ions through the solid electrolyte interphase (SEI) layer formed on the surface of the magnesium anode. Therefore, alternative anodes that are compatible with common magnesium electrolytes are applicable and useful for Mg based energy storage. The passive layer on Mg metal electrodes may be avoided using the electrode materials, based on the mechanism of intercalation, conversion and alloy formation with Mg. Recently, metallic Bi, Sb, Sn and Pb are reported as more striking negative electrode materials for Mg based batteries on the basis of reversible alloy formation and have theoretical capacities of 385 (Mg$_3$Bi$_2$), 660 (Mg$_3$Sb$_2$), 903 (Mg$_2$Sn) and 517 (Mg$_2$Pb) mAh g$^{-1}$.[18-25] But experimentally Sb and Sn delivers low capacities about 0.02 and 20% of its theoretical capacity with low coulombic efficiency and fast capacity fading. Where, in case of Pb and Bi, they deliver more than 90% of its theoretical capacity, and moreover Bi is less toxic than Pb and that of its neighbours in the periodic table, thus it has drawn quite an attention. In the following reaction (1) Bi reacts with Mg and six electron equivalents to form an alloy of magnesium bismuthide with a theoretical capacity of 385 mAh g$^{-1}$ and an average voltage of 0.25 vs Mg/Mg$^{+2}$.[18]

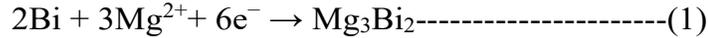

$$2Bi + 3Mg^{2+} + 6e^- \rightarrow Mg_3Bi_2 \text{----------------------(1)}$$

Arthur *et. al.*,[20] reported wide composition ranges of Bi, Sb and Bi-Sb alloys by electrodeposition. The compositions with Sb, $Bi_{0.55}Sb_{0.45}$, $Bi_{0.88}Sb_{0.12}$ and Bi initially delivered specific capacity of 16, 235, 298 and 222 mAh g$^{-1}$ respectively at 1 C rate. After several cycles, the capacity faded to 151 and 215 mAh g$^{-1}$ for $Bi_{0.55}Sb_{0.45}$ and $Bi_{0.88}Sb_{0.12}$, respectively, whereas the Bi anode showed good cycling performance with high specific capacity around 247 mAh g$^{-1}$ at the 100th cycle.[20] Similarly, Watson *et. al.*, reported that Bi is showing high electrochemical performance than Sb, due to the ionicity and Mg-Sb bonding in $Mg_3Sb_2$ which is greater than $Mg_3Bi_2$.[22] Singh *et. al.*,[19] reported Sn as high capacity anode material for Mg-ion battery with initial specific capacities of 220 and 260 mAh g$^{-1}$ and dropped to 280 and 220 mAh g$^{-1}$ at 10$^{th}$ cycle with 0.005 and 0.01 C rates, respectively. The capacities of de-magnetisation of Sn (220, 250, 210 and 200 mAh g$^{-1}$) and Bi (380, 340, 320, 290 mAh g$^{-1}$) were reported as follows with specific rates 0.005, 0.01, 0.02 and 0.05 C, respectively. These materials are showing poor rate performance and fast capacity fading. Such problems are mainly associated with the low solid-state diffusion coefficient of Mg$^{+2}$ ion and large volume expansion/compression taking place during Mg ion intercalation and deintercalation process. Recently, Shao *et. al.*,[18] reported high-performance nanostructure Bi for Mg ion battery. They prepared Bi nanotubes hydrothermally which delivered a stable discharge capacity of 303 mAh g$^{-1}$ at C/20 rate and 216 mAh g$^{-1}$ at 5C rate close to 100% coulombic efficiency up to 200 cycles. Whereas micron size Bi delivered 188 mAh g$^{-1}$ and 50 mAh g$^{-1}$ with rates of C/20 and 5C, respectively at 200$^{th}$ cycle. Bi nanotubes showed high electrochemical performance, when compared to micron sized Bi particles, due to the interconnected Nano-pores of Bi nanotubes, which favours the Mg$^{2+}$ ion insertion and disinsertion.

Using Bi alone as an electrode has its own drawbacks including sluggish magnesium insertion/extraction kinetics and electrode pulverization due to volume changes. Nanostructured bismuth is used as insertion material, which account for short diffusion lengths of Mg$^{2+}$. The result of using the Bi nanostructured insertion material as anode in magnesium energy storage systems can significantly increase charging/discharging rate and improve cycling stability. The use of RGO nanocomposite with Bi in rechargeable magnesium ion battery is expected to increase cycling stability *via* minimizing the mechanical stress during magneisation and demagneisation.[26] Presence of reduced graphene oxide layers can effectively acclimatize the large volume change without losing electric contact and considerably reducing diffusion length for Mg$^{2+}$ during alloying and dealloying.

In the present work, Bi nanoparticles anchored reduced graphene oxide was prepared by in-situ reduction of graphene oxide (GO) and Bi$^+$ ions by ethylene glycol in a basic medium under $N_2$ atmosphere using hydrazine hydrate as reducing agent. The Bi/RGO composites used as negative electrode material for Mg ion battery, which are electrochemically cycled verses Mg in non-aqueous electrolytes. These results are compared with pure Bi nanoparticles and pure RGO, which are prepared by the same synthetic procedure. In the other hand graphene not only act as an outstanding electrical conductor but also acting as buffer for the volume exposition and compression during the electrochemical Mg insertion / deinsertion and avoids the aggregation of bismuth particles.

## 2 EXPERIMENTAL SECTION

### 2.1. Chemicals

The following chemicals were obtained from different suppliers and they were used as received. Bismuth (III) nitrate pent hydrate (Aldrich), poly(vinylidene fluoride) (PVDF) (Aldrich), 1-methyl-2-pyrrolidinone (NMP) (Aldrich), 2M ethylene magnesium chloride in THF (Aldrich), graphite (Indian Carbon), acetylene black (AB) (Alfa Aesar), hydrazine hydrate (SD Fine Chemicals), potassium permanganate (SD Fine Chemicals), sodium nitrate (SD Fine Chemicals), ethylene glycol (SD Fine Chemicals), conc., $H_2SO_4$ (SD Fine Chemicals). Double distilled water was prepared in the laboratory in a two-stage quartz distillation unit.

### 2.2. Synthesis

#### 2.2.1. Preparation of graphite oxide

Graphite oxide (GtO) was prepared from graphite powder in a secure, quicker, and more efficient route called the modified Hummers' method.[27] In this typical synthesis 3 g of graphite is added to the 69 ml of concentrated $H_2SO_4$ and 1.5 g of sodium nitrate with a time interval of 20-30 min under constant stirring for 1 h at ambient temperature. The above resultant solution was cooled in an ice bath and 9 g of $KMnO_4$ was added slowly in about 15 min. The solution was placed in an ice bath to prevent overheating and explosion. Later the container was allowed to reach ambient temperature and diluted by adding 500 ml of water under stirring. Then the suspension was further oxidised with 7.5 ml of 30% $H_2O_2$. The colour of suspension changes from brown to light yellow indicating oxidation of graphite to graphite oxide. The resulting mixture was thoroughly washed with warm water and ethanol respectively. Finally filtered and dried at 50 °C overnight. The obtained graphite oxide (GtO) was further used for the preparation of Bismuth (Bi) / reduced graphene oxide (RGO) nanocomposites.

#### 2.2.2. Preparation of bismuth / reduced graphene oxide nanocomposite

Different ratios of Bi/RGO nanocomposites (Table 1) were prepared through in-situ reduction in a basic medium under $N_2$ atmosphere (Figure 1) using hydrazine hydrate as reducing agent, reported elsewhere.[28] In the typical synthesis of 50% Bi and 50% RGO nanocomposite, 250 mg of GtO powder was disseminated in 160 ml of ethylene glycol (EG) followed by ultra-sonication for 1 h 30 min. The so-formed graphene oxide colloid was mixed with 40 ml of EG solution containing 580.0 mg $Bi(NO_3)_3.5H_2O$. Subsequently, the *in-situ* reduction of graphene oxide and bismuth nitrate was carried out by adding required amount of hydrazine hydrate (15-25 ml) followed by the addition of 7.2 ml of EG solution with 1 g NaOH. The above mixture is sonicated for 10 min, transferred to a 250 ml round bottomed flask and refluxed at 110 °C for 2 h under $N_2$ gas purging. The solution was allowed to settle discarding the solvent. The obtained product was further washed thoroughly with DD water and ethanol to remove left over impurities. The product was then dried at 60 °C in a hot air oven for overnight to obtain Bi/RGO nanocomposite. The same synthesis procedure was adopted for other nanocomposites of Bi and RGO by varying the ratios of $Bi(NO_3)_3 \cdot 5H_2O$ and GtO (Table 1). From now the samples are referred as Bi100, Bi95, Bi90, Bi80, Bi70, Bi60, Bi50 and Bi0 for 100%Bi : 0% RGO, 95% Bi :05% RGO, 90% Bi : 10% RGO, 80% Bi : 20% RGO, 70% Bi : 30% RGO, 60% Bi: 40% RGO, 50% Bi : 50% RGO and 100% RGO:00%Bi, respectively.

| S. No | Sample name | Nanocomposite ratio | Amount of $Bi(NO_3)_2 \cdot 5H_2O$ (weight of Bi obtained) | Amount of GtO | Total weight of RGO/Bi |
|---|---|---|---|---|---|
| 1 | Bi100 | 100%Bi : 0%RGO | 1.160 g (500 mg) | 0 mg | 500 mg |
| 2 | Bi95 | 95% Bi : 5% RGO | 1.102 g (475 mg) | 25 mg | 500 mg |
| 3 | Bi90 | 90% Bi : 10% RGO | 1.044 g (450 mg) | 50 mg | 500 mg |
| 4 | Bi80 | 80% Bi : 20% RGO | 928.4 mg (400 mg) | 100 mg | 500 mg |
| 5 | Bi70 | 70% Bi : 30% RGO | 812.4 mg (350 mg) | 150 mg | 500 mg |
| 6 | Bi60 | 60% Bi : 40% RGO | 696.3 mg (300 mg) | 200 mg | 500 mg |
| 7 | Bi50 | 50% Bi : 50% RGO | 580.0 mg (250 mg) | 250 mg | 500 mg |
| 8 | Bi0 | 0%Bi : 100% RGO | 0 mg (0 mg) | 500 mg | 500 mg |

*Table 1. Weights of compound used in the synthesis of different ratios of nanocomposite.*

### 2.2.3. Characterization methods

Sonication assisted synthesis during all stages of preparation of RGO and RGO/Bi nanocomposites were carried out using Misonix ultra sonicator model S4000-010. The operational frequency was 20 kHz with a power output of 600 W. Sonication probe with Titanium horn (diameter = 12 mm) was dipped in the aqueous phase and sonicated for the required duration. X-ray powdered diffraction (XRD) patterns were recorded using Philips X-PertPro diffractometer at 40 kV and 30 mA using Cu Kα (λ = 01.5418 Å) radiation source. Functional group analysis was performed using 1000 Perkin Elmer FT-IR spectrometer. The morphology was examined using a Gemini Technology scanning electron microscopy (SEM) model ULTRA 55, and JEOL Co. transmission electron microscopy (TEM) model JEOL-JEM 2100F. The chemical composition and functional groups on the graphene was examined by Thermofisher Scientific X-ray photoelectron spectroscopy (XPS) using X-ray Al anode (monochromatic Kα X-rays at 1486.6 eV) as the source. The C 1S region was used as a reference and was set at 284.6 eV. Thermo-gravimetric analysis (TGA) was recorded from ambient temperature to 800 °C at a heating rate of 10 °C $min^{-1}$ under flow of $O_2$ gas using NETZSCH thermal analyzer model TG 209 FI.

All electrochemical studies were carried out in a homemade Swagelok cells. For the fabrication of electrodes, the active material (80 wt %), conductive material (Acetylene black, 10 wt %) and polyvinylidene fluoride (PVDF, 10 wt %) were mixed and ground in a mortar. Few drops of n-methyl pyrrolidone (NMP, Aldrich) were added to make it as slurry. A copper disk (12 mm diameter) was polished with successive grades of emery, degreased, etched in dilute 10% $HNO_3$, washed with detergent, rinsed with distilled water and acetone followed by drying in air.

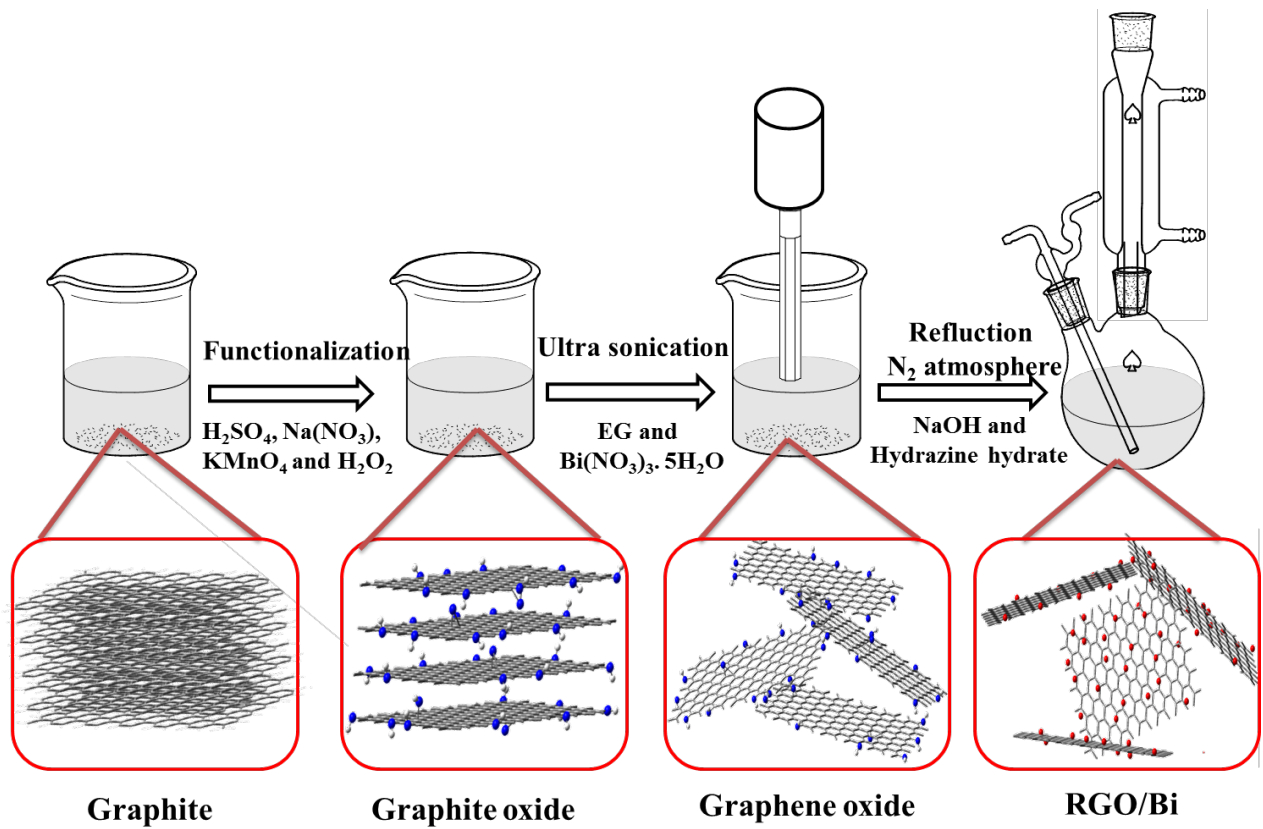

*Figure 1. Schematic diagram of synthesis process. Blue and Red colour dots represent the functional groups and Bi nanoparticles respectively.*

The slurry was coated on the pre-treated copper disk and dried at 110 °C under vacuum furnace for 12 h. Coating and drying steps were repeated till the mass of active material reached 2-5 mg $cm^{-2}$. The electrodes were weighed using a Mettler Toledo electronic balance model AB265-S/FACT with a sensitivity of 0.01 mg. Magnesium metal foil was used as the counter cum reference electrode and adsorbed glass mat (AGM) was used as a separator. Commercially available 2 M ethyl magnesium chloride ($EtMgCl_2$) in terahydrofuran (THF) was used as an electrolyte. The cells were assembled in an argon filled glove box MBraun model UNILAB. The cells were galvanostatically cycled in the potential range from 0.005 to 0.6 V *vs*. $Mg/Mg^{2+}$ at different current densities. Cyclic voltammetry, galvanostatic charge-discharge cycling and rate capability experiments were carried out by using a Biologic SA multichannel potentiostat/galvanostat model VMP3 and Bitrode battery cycling unit. All electrochemical studies were conducted in an air-conditioned room at 22±1 ºC.

## 3 Results and Discussion

The XRD pattern of graphite, GtO, RGO and RGO/Bi nanocomposite samples are shown in Figure 2a and 2b, respectively. The XRD pattern of the sample graphite (Figure 2a curve

i) is characterized by the strong (002) reflection at 2θ=26.3° corresponding to hexagonal graphitic structure. The inter layer distance 3.38 Å is obtained for (002) reflection which is comparable with the literature value. [29] After the conversion of graphite to graphite oxide, there is a shift in the (002) reflection (2θ=10.4°), this value corresponds to an interlayer distance of 7.98 Å indicating a remarkable expansion in graphite layers due to the formation of oxygen functional groups on the graphene sheets.

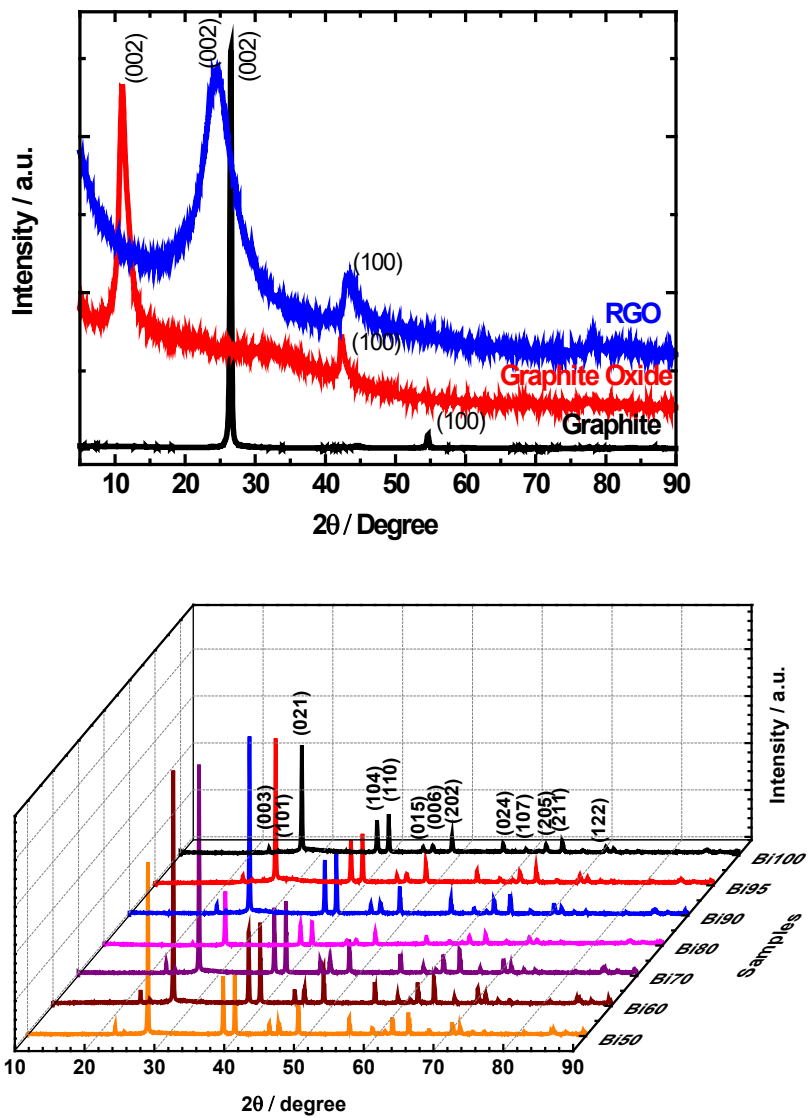

Figure 2. XRD patterns of samples (a) graphite (i), GtO (ii) and RGO (iii) and (b) Bi100 (i), Bi95 (ii), Bi90 (iii), Bi80 (iv), Bi70 (v), Bi60 (vi) and Bi50 (vii).

Reduction of graphite oxide results in shift of the (002) reflection to 2θ=23.9° with low intensity broad curve, due to the interlayer stacking of graphene sheets. The broad, weak (002) reflection is also attributed to small size (<1 μm) and a short order domain (or) turbostatic arrangement of reduced graphene oxide stacked sheets.[33] The XRD patterns of Bi/RGO nanocomposite samples are shown in Figure 2b. From the XRD pattern of the above composite it is evident that the diffraction peak of the disordered stack of graphene sheets in the composites at 2θ of 20-30° disappeared due to the less agglomeration of graphene sheets. The remaining diffraction peaks of Bi are indexed to rhombohedra phase, which are consistent with the literature values (JCPDS No. 05-0519). [30-32] No impurities were detected in the pattern, which indicates that pure Bismuth was formed without any impurities under present synthetic condition. The average crystal size of Bi particle is 100-200 nm calculated by Scherrer formula.

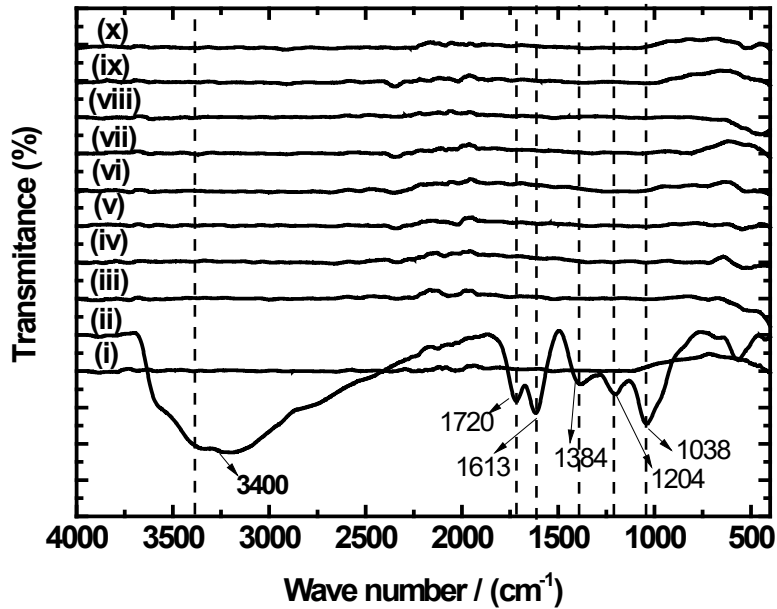

*Figure 3. FTIR spectra of graphite (i), GtO (ii), RGO (iii), Bi100 (iv), Bi95 (v), Bi90 (vi), Bi80 (vii), Bi70 (viii), Bi60 (ix), and Bi50 (x).*

Figure 3 represents the FT-IR spectra of the samples. A broad and intensive absorption band at 3400 cm$^{-1}$ is originated from the O-H bond stretching vibration on the graphene oxide surface. The bands at 1723 and 1605 cm$^{-1}$ represent the vibrations from the C=O and C=C bond stretch, respectively.[33-35] The bond stretching vibration peaks of C-O (epoxy), C-O (alkoxy) and the deformation peak of O-H are as follows respectively 1204, 1038 and 1384 cm$^{-1}$. In the spectra of RGO and RGO/Bi nanocomposites, the disappearance of absorption bands at 3250 and 1723 cm$^{-1}$ (C=O) confirms the reduction of graphite oxide to reduced graphene oxide.

The Raman spectra of all the samples are as shown in Figure 4. The D and G bands at 1347 and 1580 cm$^{-1}$ corresponds to disorder and local defects, due to the breathing k-point phonon of A1g symmetry and the E2g phonon of C sp$^2$ atom, respectively. The D- band is absent in the Raman spectra of defect free samples. When metal nanoparticles were deposited on graphene, the intensity of the D-band increased relative to the G-band and both D and G-bands shifted to higher wave numbers. A relationship between ionization energy of the metal and shift in the position of G-band was proposed, *i.e.*, the magnitude of the G-band shift decreases with an increase in ionization energy of the metal.[33-35] The intensity ratio between D and G bands for RGO (Figure 4 curve iii) and Bi/RGO nanocomposites (Figure 4 curve iv to viii) are about 1. Furthermore, there is a shift of G band from 1579 cm$^{-1}$ in RGO to 1582 cm$^{-1}$ in Bi/RGO, thus the shift in G-band is 3.0 cm$^{-1}$. These features of Raman spectra confirm the electronic interactions of the metal particles with graphene.

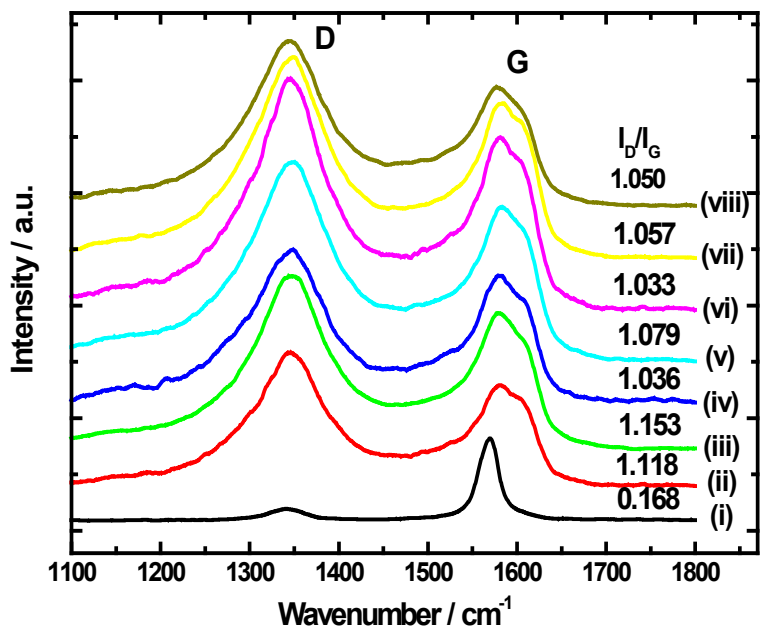

Figure 4. Raman spectra of graphite (i), GtO (ii), RGO (iii), Bi50 (iv), Bi60 (v), Bi70 (vi), Bi80 (vii), Bi90 (viii), Bi95 (ix) and Bi100 (x).

The XPS spectra samples are shown in Figure 5 for GtO, RGO and Bi/RGO nanocomposites, respectively. Figure 5 shows the deconvoluted XPS C 1s spectra fitted with Gaussian-Lorentzian peak shape after performing a Shirley background correction. The deconvoluted XPS C 1s spectra shows four characteristic peaks at 284.5, 285.3, 286.6 and 288.8 eV for sp2 hybridized C, sp3 hybridized C, epoxy or carboxy and carbonyl group moieties, respectively. [49-51] After reduction of graphite to RGO (Figure 5b), sp3 hybridized carbon (C-C) peak intensity decreased and sp2 hybridized carbon (C=C) peak intensity increased. The intense ratio of C=C/C-C increased from 0.23 to 0.33 for GtO (Figure 5a) to RGO (Figure 5b). The intensities of epoxy or carboxy and carbonyl groups at 286.7 and 288.8 eV were remarkably reduced after reduction in all RGO and RGO/Bi nanocomposite samples (Figure 5b-h). The thermogravimetric curves of graphite, GtO, RGO and RGO/Bi nanocomposites are shown in Figure 6. There is no detectable weight loss for graphite (Figure6, curve i) in the temperature range of 600 °C, but 43 % weight loss is observed between 600 to 900 °C due to the combustion of graphite. [33-35]

Figure 6, Curve ii, for sample GtO, ~15% weight loss up to 110 °C may be resulting due to removal of water molecule trapped between the graphene oxide sheets; ~20% weight loss at 300 °C could possibly because of the pyrolysis of functional groups, [35] ~50% weight loss at 600 °C can be corresponded to combustion of carbon skeleton of GtO. Curve iii for sample RGO shows good thermal stability and it is showing continuous weight loss of about 60%, due to the elimination of remaining functional groups.

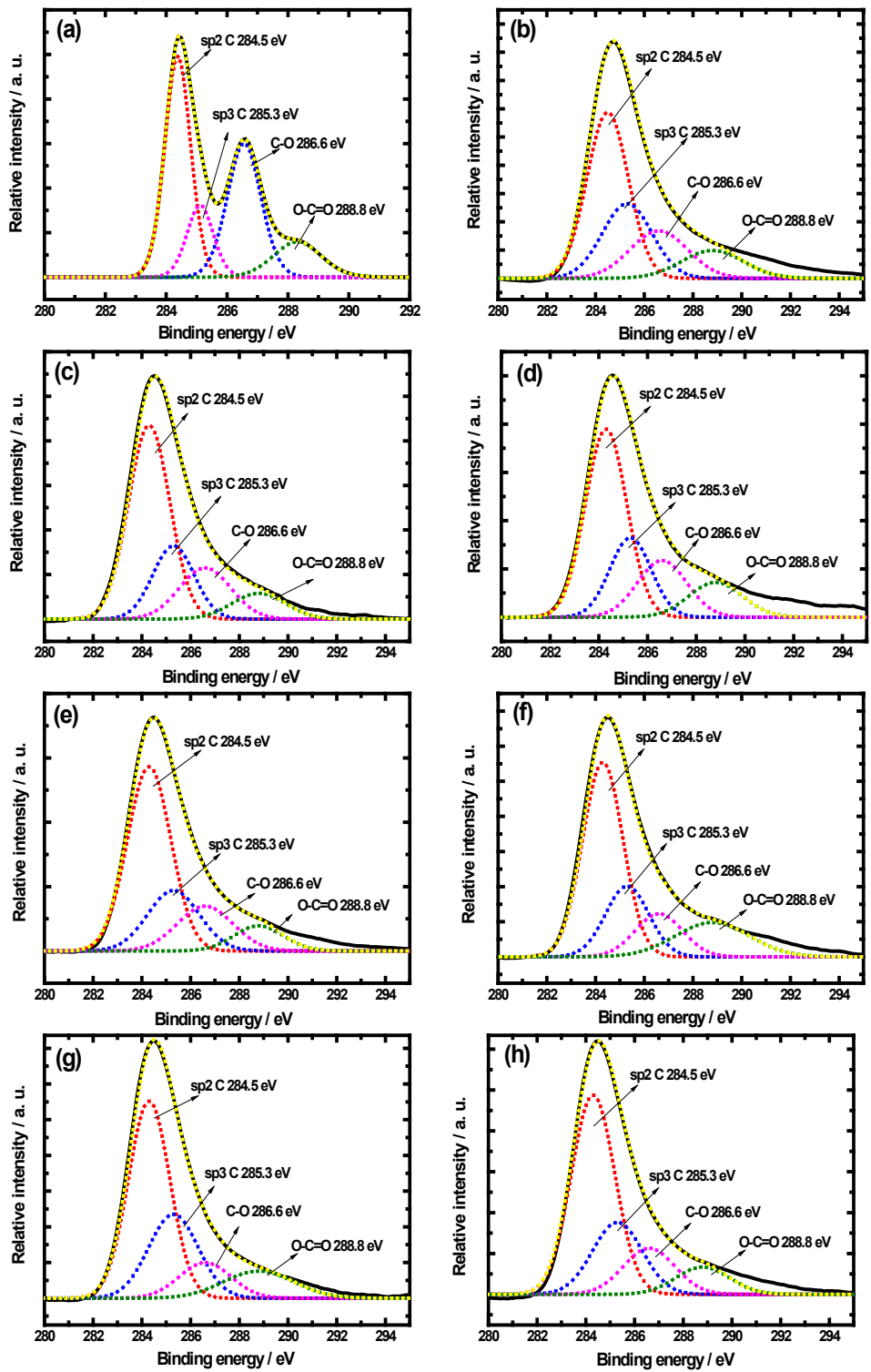

*Figure 5. XPS spectra of samples (a) GtO, (b) RGO, (c) Bi95, (d) Bi90 (e) Bi80, (f) Bi70, (g) Bi60, and (h) Bi50.*

About 31% weight loss at 450 ºC is observed for sample Bi60 (Figure 6 curve iv) and 4% weight loss is observed for Bi95 samples (Figure 6, curve v), due to the removal of RGO and its functional groups from the nanocomposites. Bi100 showing (Figure 6, curve vi) only 1% loss which indicates the absence of RGO, and Bi is seemed to be quite thermally stable.

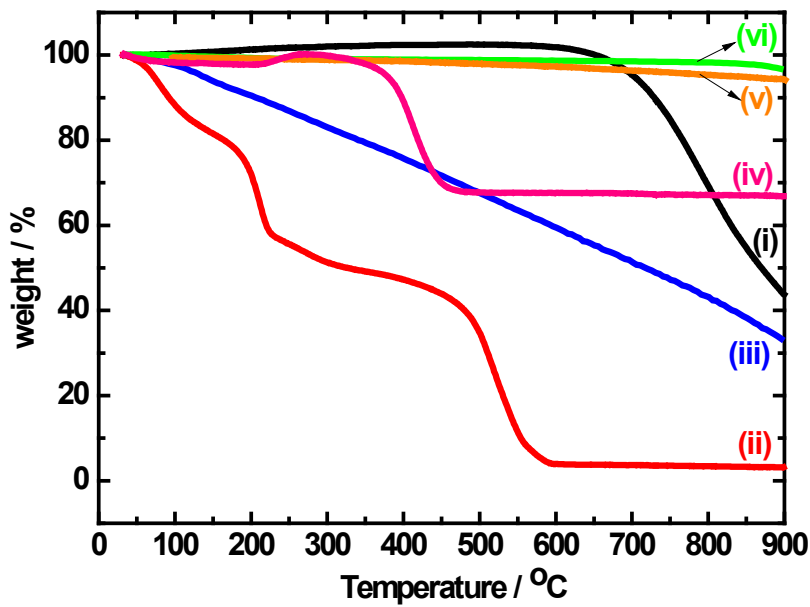

*Figure 6. TGA curves of the samples, graphite (i), GtO (ii), RGO (iii), Bi60 (iv), Bi95 (v) and Bi100 (vi).*

Figure 7 and 8 show the SEM and TEM morphological studies of all the samples, respectively. Morphology changes from Fig 7a to c indicates the formation of RGO by clear exfoliation and separation of few layers of graphene oxide sheets from graphite by introducing the functional groups. Figure 7d shows the well-grown Bi nanoparticles with an average size of 500 nm. Fig. 7e to j represents the anchoring of Bi nanoparticles in graphene sheets and it is noticed that the percentage of RGO is increased in the nanocomposite from samples Bi95 to Bi50. The TEM images (Fig. 8) reveal that single layers of graphene sheets as well high density Bi metal nanoparticles were uniformly dispersed on the reduced graphene oxide sheets with negligible aggregation. No free Bi nanoparticles were observed outside the graphene sheet even after sonication of the sample for preparation of TEM characterization. This confirmed the strong interaction between RGO and Bi nanoparticles. The transparent and thin graphene sheets can be distinguished and the calculated lattice spacing is about 0.36 and 0.35 nm from the lattice fringes of HRTEM images of RGO and Bi particles, respectively, which is corresponding to the (002) plane for RGO and (101) plane of the rhombohedral structure of Bi nanoparticle (Fig. 8 (i) and (j)). Figure 8 (k) represents EDS spectra of Bi60 sample and it revels the existing of Bi and carbon and oxygen (from the unreduced graphene). Moreover by observing the interface between Bi and RGO, it turned out that Bi nanoparticles were well anchored on the surface of the graphene sheets.

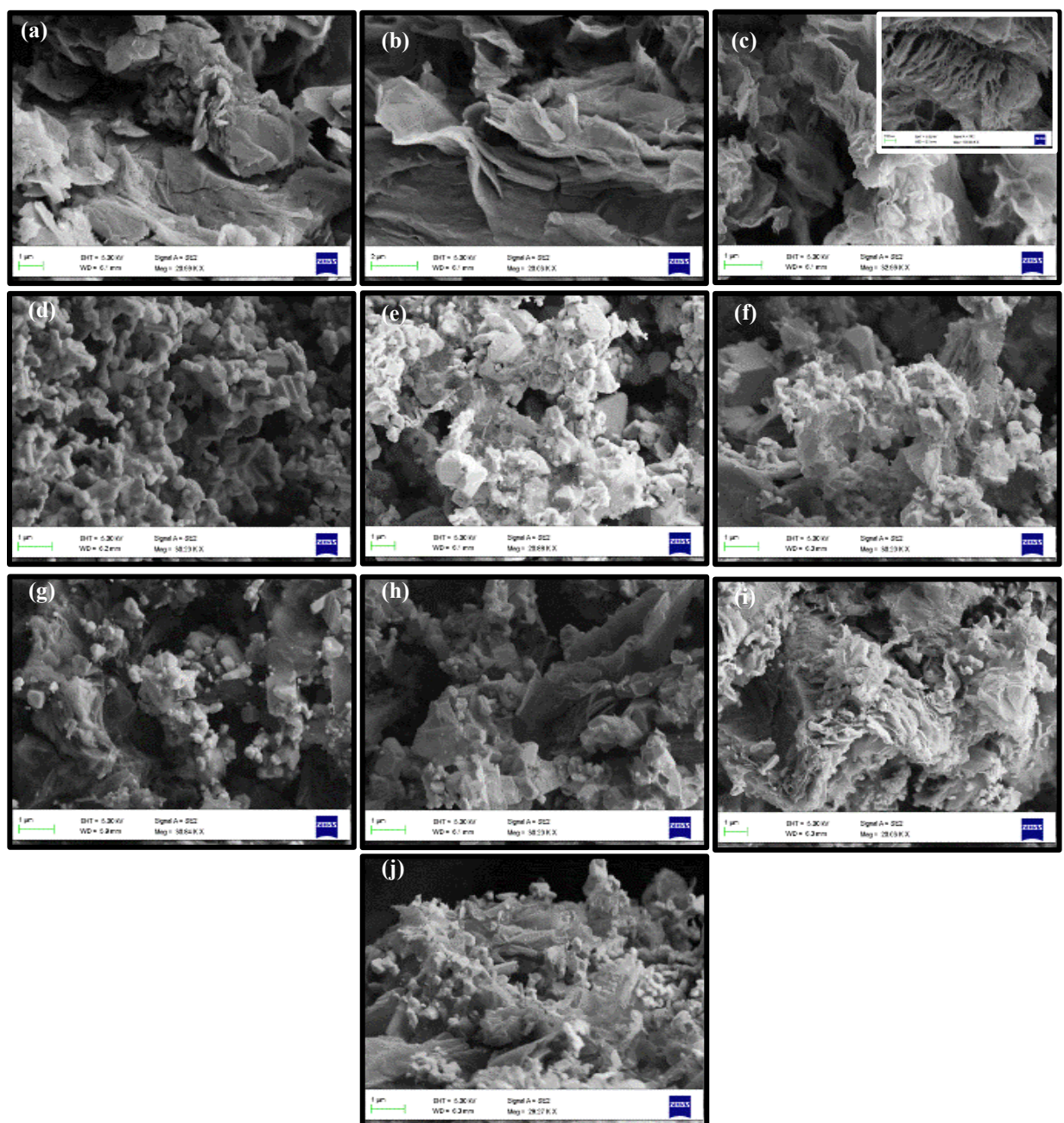

*Figure 7. SEM images of (a) Graphite, (b) GtO, (c) RGO, (d) Bi100, (e) Bi95, (f) Bi90, (g) Bi80, (h) Bi70, (i) Bi60 and (j) Bi50.*

The electrochemical performance of nanostructured materials is evaluated by cyclic voltammetry. Figure 9 shows the voltammograms of Bi0, Bi100, Bi50, Bi60, Bi70 and Bi95 samples recorded at 0.05 mV s$^{-1}$ in the potential window of 0.005 to 0.6 V vs Mg/Mg$^{+2}$. In the voltammogram, Mg insertion (magnetisation) takes place around 0 to 0.25 V and disinsertion (demagnetisation) takes place around 0.3 to 0.56 V.

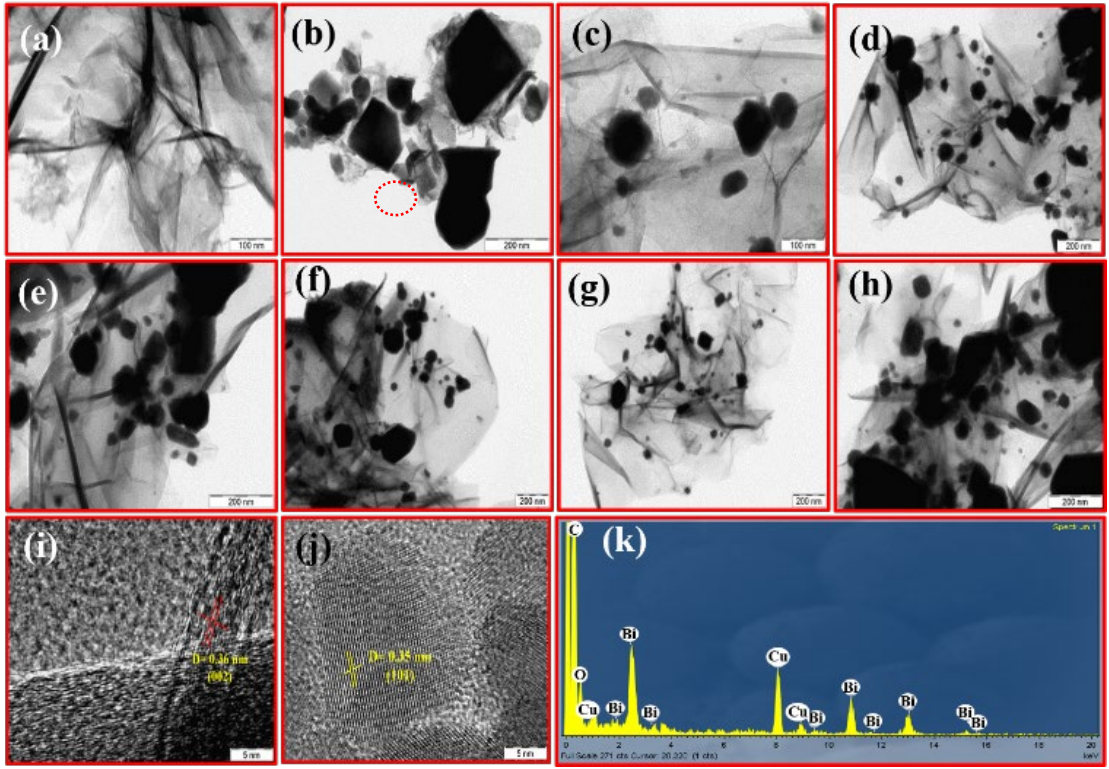

*Figure 8. TEM images of (a) RGO, (b) Bi100, (c) Bi95, (d) Bi90, (e) Bi80, (f) Bi70, (g) Bi60 and (h) Bi50. (i) and (j) represents HRTEM of RGO and Bi samples, and (k) represents EDS spectra of Bi60 sample (the spectrum acquired is marked with red circle in Fig. 8g), respectively.*

The cyclic voltammogram of RGO (Fig. 9a) shows low current redox peaks, which does not contribute much in the Mg insertion and disinsertion. In the Figure 9 b-f, the reversible redox peaks are attributed to the reversible formation of magnesium bismuthide as indicated by the following equation 2:

$$3Mg^{+2} + 2Bi + 6e^- \rightleftarrows Mg_3Bi_2 \text{------------------ (2)}$$

From the second cycle onwards, the redox peaks shows sharp and narrow peaks, indicating the reversible and fast Mg ion insertion and disinsertion.

The charge discharge curves of all the nanocomposite samples shown in Figure 10 are in the voltage range between 0.005 and 0.60V vs Mg/Mg$^{+2}$ at specific current around 25 mA g$^{-1}$. During the first charging process, from open circuit voltage to 0.005 V vs Mg/Mg$^{+2}$, one flat plateau region is observed at 0.2 V, which is attributed to the Mg$^{+2}$ ion insertion to the electrode material. Wherein during discharging process, plateau region observed at around 0.3 V vs Mg/Mg$^{+2}$ is due to the Mg ion disinsertion. The charging capacity of Bi100 sample in the first cycle is 392 mAh g$^{-1}$ and corresponding discharge capacity is 326 mAh g$^{-1}$ at current density of 19.9 mA g$^{-1}$ with coulombic efficiency about 87%. In the second cycle, the charge and discharge capacities are 332 and 326 mAh g$^{-1}$, respectively.

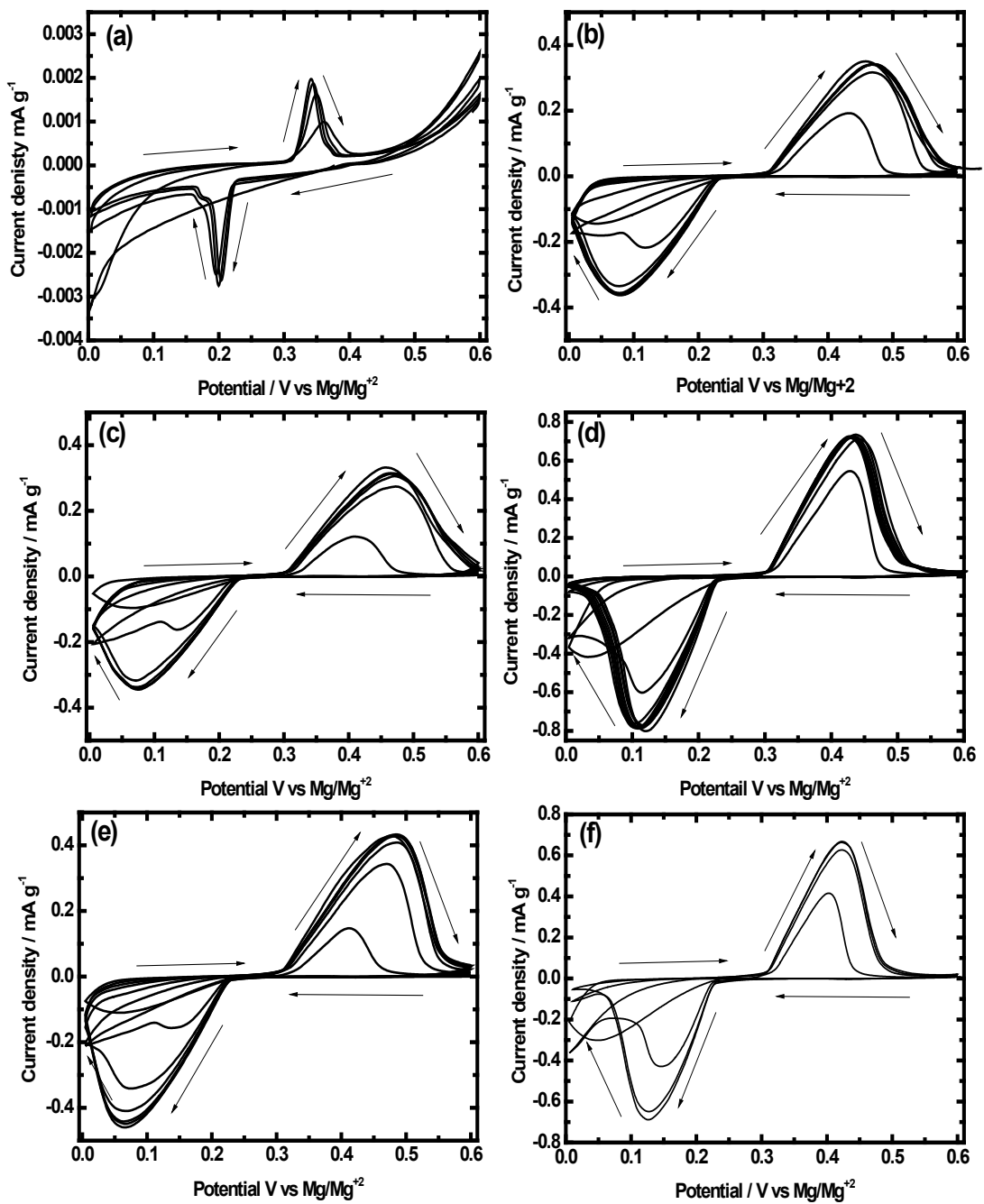

*Figure 9. Cyclic voltammograms of samples (a) RGO (Bi0), (b) Bi100, (c) Bi95, (d) Bi70, (e) Bi60 and (f) Bi50.*

The columbic efficiency, thus improved in the second and subsequent cycles close to 95%. The charge plateau for the second cycle increased to 0.25 V vs Mg/Mg$^{+2}$. The irreversible capacity loss observed in the first cycle is 50 mAh g$^{-1}$, corresponds to phase stabilization of electrode material. Similar observations are made for other nanocomposite samples Bi50, Bi60, Bi70, Bi80 and Bi95, the first cycle discharge capacity values are 350, 414, 363, 338 and 333 mAh g$^{-1}$ and corresponding charge capacities are 505, 547, 461, 433 and 406 mAh g$^{-1}$ with coulombic efficiencies up to 69, 76, 79, 78 and 82%, respectively. The columbic efficiency is thus improved in the second and subsequent cycles up to more than 97% for all the nanocomposites.

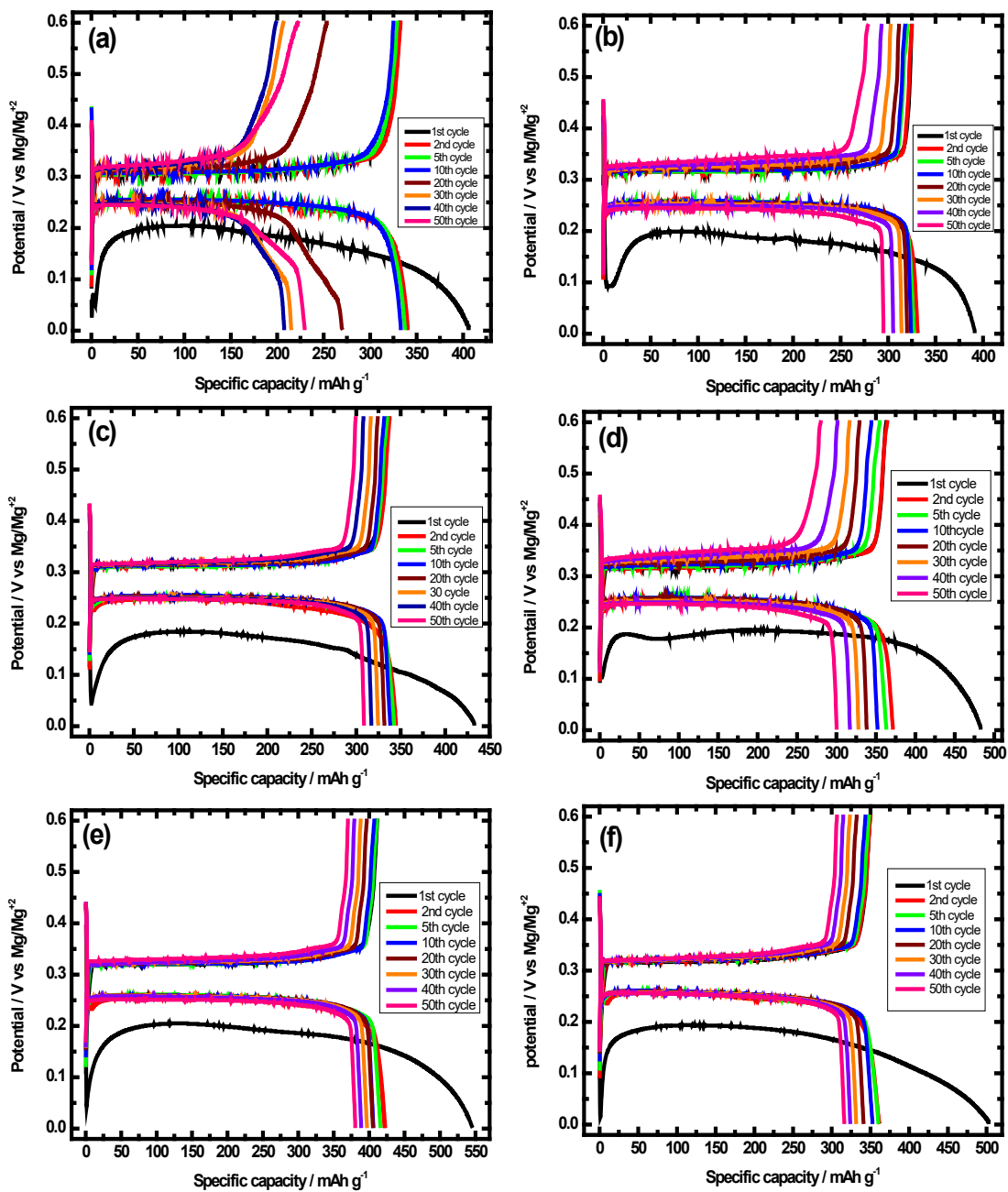

*Figure 10. Charge/discharge profiles of samples (a) Bi100, (b) Bi95, (c) Bi80, (d) Bi70, (e) Bi60 and (f) Bi50.*

It is observed that the cycle performance of Bi60 sample is superior to the other composite samples. The mechanism is still unclear, the charge/discharge profiles (Fig. 10) and cyclic voltammogram (Fig. 9), the initial cycles are different from the subsequent cycles, due to the activation associated with the formation of solid electrolyte interface. Further studies are needed to determine the precise mechanisms of activation during the first cycle.

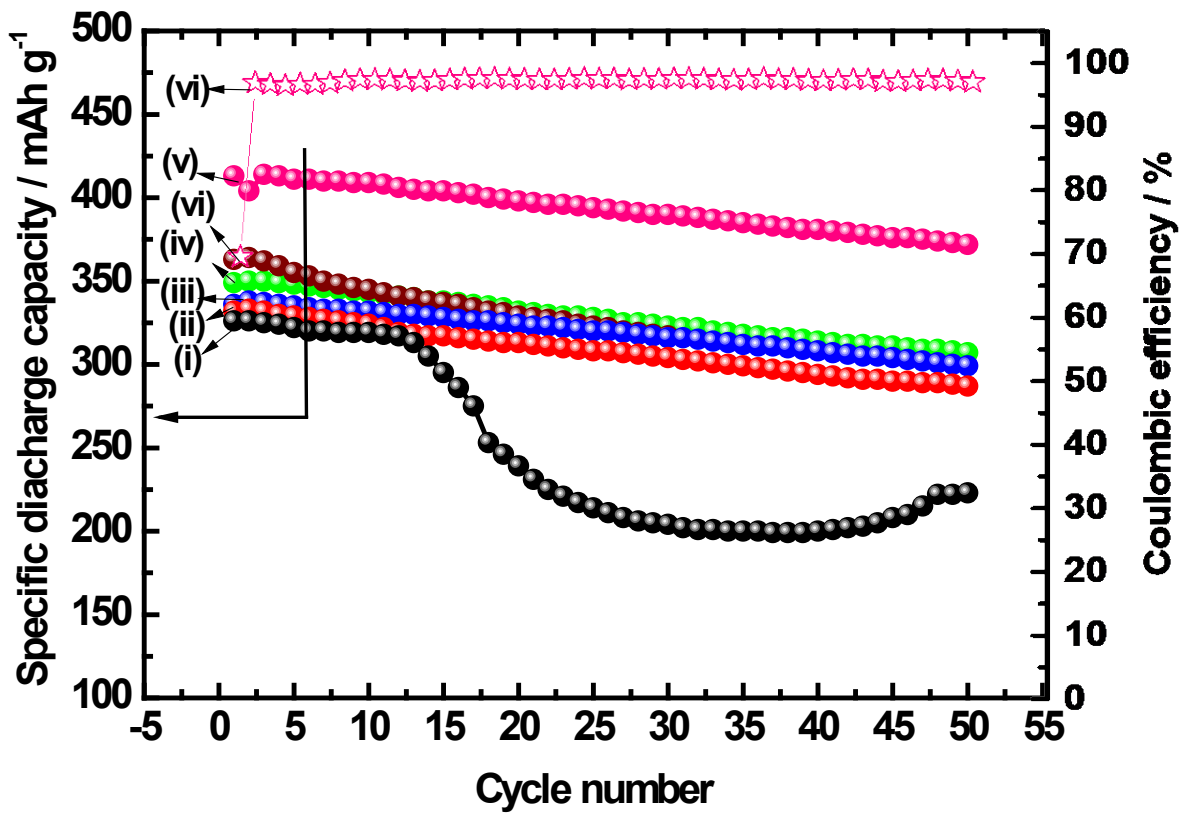

*Figure 11. Cycle life stability and columbic efficiency (Bi60) of samples are represented for 50 cycles: (i) Bi100, (ii) Bi95, (iii) Bi80, (iv) Bi70, (v) Bi60 and (vi) Bi50.*

Figure 11 shows the cycling stability of nanocomposite samples subjected to galvanostatic charge/discharge cycling at their respective specific current density in the voltage range between 0.005 and 0.6 V vs Mg/Mg$^{+2}$ for 50 cycles. For the Bi100 sample, the initial and the 50th cycle discharge capacity values of 326 and 285 mAh g$^{-1}$ are obtained at the current density of 20 mA g$^{-1}$. Samples Bi50, Bi60, Bi70, Bi80 and Bi95 delivered the initial discharge capacity values of 350, 414, 363, 338, 333 mAh g$^{-1}$ and the values at the 50th cycles are 307, 372, 304, 299, 222 mAh g$^{-1}$ at specific currents 33, 39, 24, 29 and 15 mA g$^{-1}$, respectively. The nanocomposite of Bi60 performs better than the other composite samples.

In order to study the rate capability, nanocomposite samples are subjected to charge/discharge cycling at different current densities in the range of 20 to 700 mA g$^{-1}$ (Figure 12). The discharge capacity values tends to be decreased, delivering about 326,

321, 301, 274, 57 and 1 mAh $g^{-1}$ at current densities 20 (C/20), 32 (C/12), 100 (C/4), 200 (C/2), 500 (1.3C) and 700 (2C) mA $g^{-1}$, respectively for Bi100. Whereas the Bi60 samples delivered the discharge capacities of 414, 381, 372, 354, 295, 238, 363 and 362 mAh $g^{-1}$ at current densities 39 (C/10), 53 (C/7), 100 (C/4), 200 (C/2), 500 (1.3C), 700 (2C) mA $g^{-1}$, respectively.

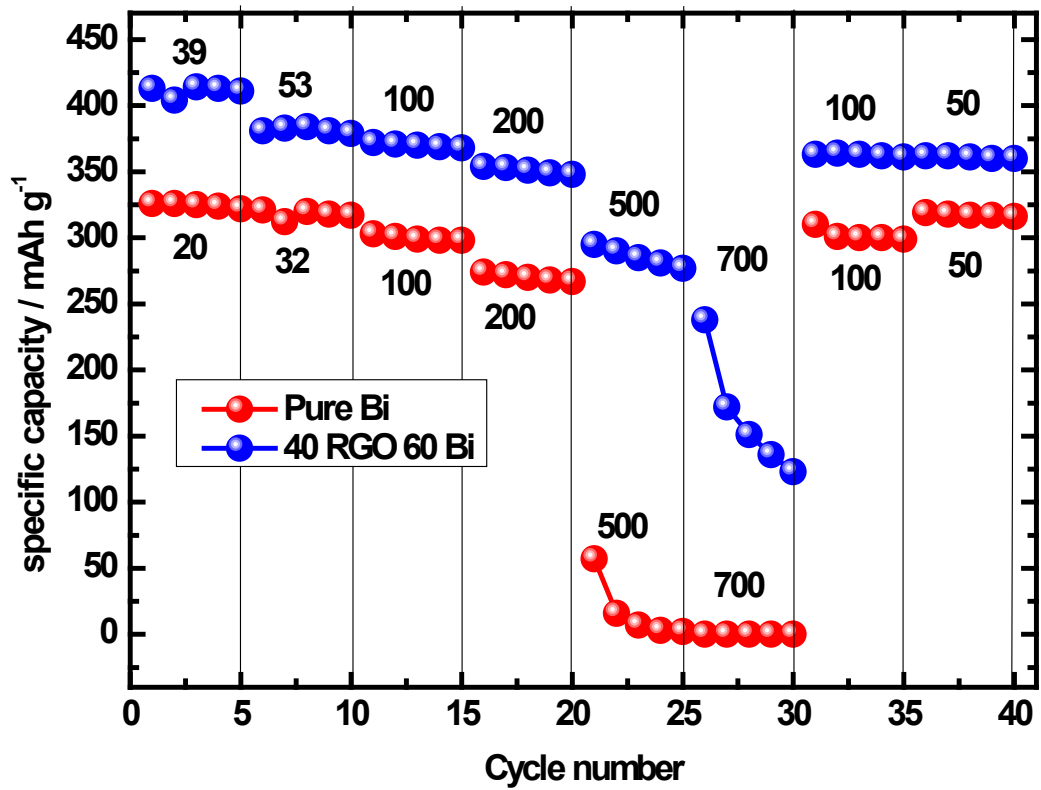

*Figure 12. Rate capability of sample Bi100 and Bi60 at different current densities are indicated in mA g-1.*

The C-rate are calculated based on the theoretical specific capacity of 385 ($Mg_3Bi_2$). After cycling at high rates (high current density 700 mA $g^{-1}$, the specific capacity returns to low current densities (50 and 100 mA $g^{-1}$), the samples Bi100 and Bi60 have exhibited the discharge capacities about 300 and 372 mAh $g^{-1}$, respectively, which indicates that the samples are highly robust and stable.

## 4 Conclusions

RGO/Bi nanocomposite was synthesized by in-situ reduction in solvothermal condition at 100 oC under $N_2$ atmosphere and studied as an anode material for Mg-ion battery. The formation of RGO, Bi and RGO/Bi nanocomposite were confirmed by physical characterizations. The 40% RGO and 60% Bi showed good electrochemical performance in terms of high discharge capacity with good cycle life and rate capability. The observed improvement in electrochemical performance might be accredited by graphene oxide layers, which increases the electronic conductivity and accommodate significant volume changes. 40% RGO/60% Bi sample delivered discharge capacity as high as 372 mAh $g^{-1}$ at a specific current of 39 mA g-1 in the 50th cycle and possess high rate capability with discharge capacity of 238 mAh g$^{-1}$ at specific current 700 mA g$^{-1}$. Apparently, reduced graphene oxide (RGO)/bismuth (Bi) nanocomposite appeared to be a promising high capacity anode for magnesium ion batteries with longer cycle life and high-rate performance.

**Conflict of Interest**

Authors declare that there is no conflict of interest.

**Acknowledgements**

TRP and GV acknowledge the financial support from Indian Institute of Science, Bangalore under Research Associate Fellowship and Bachelor of Science (Research) Programme. As well acknowledge SSK for the financial support from the University Grant Commission (UGC), Government of India, under Dr. D.S. Kothari fellowship program [Ref. No. F.4-2/2006(BSR)/13-626/2012(BSR)] UGC for the financial support and thank the CeNSE, IISc for providing characterisation facilities.

**Notes and references**

## Abstract

Herein, *in-situ* reduction of bismuth and graphene oxide was performed by a solvothermal method under a nitrogen atmosphere, and the resulting Bi/RGO nanocomposites were used as an anode material for Mg-ion batteries. The nanocomposite of 60% Bi : 40% RGO is a beneficial anode material, delivering a discharge capacity as high as 413 and 372 mAh/g at the specific current of 39 mA/g in the $1^{st}$ and $50^{th}$ cycles, respectively. In addition, it shows high-rate capability with the discharge capacities of 381, 372, 354, 295, and 238 mAh/g at the specific currents of 53, 100, 200, 500, and 700 mA/g, respectively. The better electrochemical performance of the nanocomposite is due to improvement in the electronic conductivity and significant reduction of volume changes during electrochemical cycling. This study demonstrates the bismuth (Bi)/reduced graphene oxide (RGO) nanocomposite as a promising high-capacity anode for magnesium-ion batteries with longer life cycle and high-rate performance.